\documentclass[12pt]{article}

\usepackage{epsfig}

\newcommand{\be}{\begin{eqnarray}}
\newcommand{\ee}{\end{eqnarray}}

\def\lsim{\mathrel{\rlap{\lower4pt\hbox{\hskip1pt$\sim$}}
    \raise1pt\hbox{$<$}}}               
\def\gsim{\mathrel{\rlap{\lower4pt\hbox{\hskip1pt$\sim$}}
    \raise1pt\hbox{$>$}}}               

\def\NPB{{\em Nucl. Phys.} B~}
\def\PLB{{\em Phys. Lett.} B~}
\def\PRL{{\em Phys. Rev. Lett.}~}
\def\PRC{{\em Phys. Rev.} C~}
\def\PRD{{\em Phys. Rev.} D~}

\begin{document}

\rightline{\Large{Preprint RM3-TH/03-8}}

\vspace{1.0cm}

\begin{center}

\LARGE{Longitudinal structure function of the proton at low momentum transfer and extended constituents\footnote{{\bf To appear in Physics Letters B.}}}

\vspace{1.0cm}

\large{Silvano Simula}

\vspace{0.5cm}

\normalsize{Istituto Nazionale di Fisica Nucleare, Sezione di Roma III,\\ Via della Vasca Navale 84, I-00146, Roma, Italy}

\end{center}

\vspace{1.0cm}

\begin{abstract}

\indent Recent data on the (inelastic) Nachtmann moments of the unpolarized proton structure function $F_2^p$, obtained at low momentum transfer with the $CLAS$ detector at Jefferson Lab, have been interpreted in terms of the dominance of the elastic coupling of the virtual photon with extended substructures inside the proton.  Adopting the same constituent form factors and the same light-front wave function describing the motion of the constituents in the proton, the (inelastic) Nachtmann moments of the longitudinal proton structure function $F_L^p$ are calculated (without further parameters) for values of the squared four-momentum transfer $0.2 \lsim Q^2 ~ (GeV/c)^2 \lsim 2$. The different role played by the Pauli form factor of the constituents in the transverse and longitudinal channels is illustrated. Our predictions, including an estimate of the theoretical uncertainties, may be checked against the forthcoming results of the experiment $E94110$ at Jefferson Lab. A positive comparison with the new data may provide compelling evidence that constituent quarks are intermediate substructures between the hadrons and the current quarks and gluons of $QCD$.

\end{abstract}

\vspace{1.5cm}


\vspace{0.5cm}


\newpage

\pagestyle{plain}

\section{Introduction \label{sec:introduction}}

\indent The idea to assume extended constituents for constructing a model of the Deep Inelastic Scattering ($DIS$) structure functions of hadrons dates back to the 1970s \cite{Altarelli}. The basic physics assumption is that hadrons are made of a finite number of constituent quarks ($CQ$'s) having a partonic structure. At large values of the squared four-momentum transfer $Q^2$ the virtual boson probes one $CQ$ and sees its partonic structure. The latter depends only on short-distance, perturbative physics, which is independent of the particular hadron, while the motion of the $CQ$'s inside the hadron reflects the non-perturbative physics, which depends on the particular hadron. Thus, the parton motion in a hadron is assumed to have two different stages: the first one is the (hadron-independent) motion inside the $CQ$, while the second stage is the $CQ$ motion in the hadron. We will refer here after to the model of Ref.~\cite{Altarelli} as the two-stage model. In Ref.~\cite{PSR} the two-stage model was extended to values of $Q^2$ around and below the scale of chiral symmetry breaking, $\Lambda_{\chi}$, and above the $QCD$ confinement scale, $\Lambda_{QCD}$, i.e. $0.1 \div 0.2 \lsim Q^2 ~ (GeV/c)^2 \lsim 1 \div 2$. In what follows we will refer to the model of Ref.~\cite{PSR} as the generalized two-stage model.

\indent Recently the inclusive electron-proton cross section has been measured in Hall B at Jefferson Lab ($JLab$) using the $CLAS$ spectrometer \cite{CLAS}. The measurements have been performed in the nucleon resonance regions ($W < 2.5 ~ GeV$) for values of $Q^2$ below $\approx 4.5 ~ (GeV/c)^2$. One of the most relevant feature of such measurements is that the $CLAS$ large acceptance has allowed to determine the cross section in a wide two-dimensional range of values of $Q^2$ and $x = Q^2 / 2M \nu$. This has made it possible to extract the proton structure function $F_2^p(x, Q^2)$ and to directly integrate all the existing world data at fixed $Q^2$ over the whole significant $x$-range for the determination of the transverse proton moments $M_n^{(T)}(Q^2)$ with order $n = 2, 4, 6, 8$. More precisely, the experimental moments constructed in Ref.~\cite{CLAS} are inelastic Nachtmann moments, defined as \cite{Nachtmann}
 \be
    M_n^{(T)}(Q^2) \equiv \int_0^{x_{\pi}} dx {\xi^{n+1} \over x^3} {3 + 3 
    (n + 1) r + n (n + 2) r^2 \over (n + 2) (n + 3)} ~ F_2^p(x, Q^2)
    \label{eq:MTn}
 \ee
where $\xi \equiv 2 x / (1 + r)$ is the Nachtmann variable, $r \equiv \sqrt{1 + 4 M^2 x^2 / Q^2}$ and $x_{\pi} = Q^2 / [Q^2 + (M + m_{\pi})^2 - M^2]$ is the pion threshold, with $M (m_{\pi})$ being the nucleon(pion) mass. In Eq.~(\ref{eq:MTn}) the integration over $x$ is limited up to the pion threshold in order to exclude the contribution of the proton elastic peak; therefore in what follows we will denote by $M_n^{(T)}(Q^2)$ only the contribution of the inelastic channels.

\indent A possible interpretation of the experimental results of Ref.~\cite{CLAS} has been proposed in Ref.~\cite{PSR}. There, it was shown that the data of Ref.~\cite{CLAS} exhibit a new type of scaling behavior expected within the generalized two-stage model, and that the resulting scaling function can be interpreted as (the square of) a constituent quark ($CQ$) form factor. The main conclusions of Ref.~\cite{PSR} are that: ~ i) at low momentum transfer the inclusive proton structure function $F_2^p(x, Q^2)$ originates mainly from the elastic coupling with extended objects (the $CQ$'s) inside the proton, and ~ ii) the $CQ$ size is $\approx 0.2 \div 0.3 ~ fm$.

\indent The aim of this letter is to apply the generalized two-stage model of Ref.~\cite{PSR} to the calculation of the Nachtmann moments of the longitudinal structure function, defined as \cite{Nachtmann}
 \be
    M_n^{(L)}(Q^2) \equiv \int_0^{x_{\pi}} dx {\xi^{n+1} \over x^3} \left[ 
    F_L^p(x, Q^2) + {4 M^2 x^2 \over Q^2} F_2^p(x, Q^2) {(n + 1) ~ \xi / x - 
    2 (n + 2) \over (n + 2) (n + 3) } \right]
    \label{eq:MLn}
 \ee
where
 \be
    F_L^p(x, Q^2) \equiv F_2^p(x, Q^2) ~ (1 + 4 M^2 x^2 / Q^2) - 2x F_1^p(x, 
    Q^2)
    \label{eq:FLp}
 \ee
Again, in what follows we will denote by $M_n^{(L)}(Q^2)$ only the contribution of the inelastic channels. In performing our theoretical estimate of the longitudinal moments neither the model parameters of Ref.~\cite{PSR} are changed nor further parameters are introduced. Our predictions, including an estimate of the theoretical uncertainties, may be tested against the forthcoming results of the $JLab$ experiment $E94110$ completed recently in Hall C. The latter will provide longitudinal data in the $x$-space and not directly in the moment space. However, thanks to the factor $\xi^{n + 1} / x^3$ ($< x^{n - 2}$) appearing in the r.h.s. of Eq.~(\ref{eq:MLn}), the moments with order $n \geq 4$ for $Q^2 \lsim 2 ~ (GeV/c)^2$ are mainly governed by kinematical regions accessible by the $E94110$ experiment\footnote{We have explicitly cheked this point using the parameterizations of the proton structure functions of Ref.~\cite{Ricco}. At $Q^2 = 2 ~ (GeV/c)^2$ the contribution of the resonance region ($W \leq 2 ~ GeV$) to the integral (\ref{eq:MLn}) turns out to be $\approx 70\%, 88\%, 96\%$ and $99\%$ for $n = 4, 6, 8$ and $10$, respectively. At lower values of $Q^2$ the above percentages increase for each value of the order $n$.}. We stress that a positive comparison with the new $JLab$ data may give a strong confirmation of the generalized two-stage model of Ref.~\cite{PSR} and may provide compelling evidence that $CQ$'s are intermediate substructures between the hadrons and the current quarks and gluons of $QCD$.

\indent The plan of this letter is as follows. In Section \ref{sec:transverse} we recall the basic ingredients of the generalized two-stage model of Ref.~\cite{PSR}, which is then applied to the calculation of the longitudinal Nachtmann moments in Section \ref{sec:longitudinal}. The different role played by the $CQ$ Pauli form factor in the transverse and longitudinal channels is illustrated, and our predictions, including an estimate of the theoretical uncertainties, are presented for $0.2 \lsim Q^2 ~ (GeV/c)^2 \lsim 2$. Conclusions and outlooks are collected in Section \ref{sec:conclusions}.

\section{Generalized two-stage model and the transverse channel \label{sec:transverse}}

\indent The two-stage model, proposed for the interpretation of $DIS$ data, can be extended away from $DIS$ kinematics in a simple way. Indeed, it can be argued \cite{PSR} that in a $DIS$ experiment at high values of $Q^2$ the internal structure of the $CQ$'s is probed, whereas for sufficiently low values of $Q^2$ such a structure cannot be resolved any more. Generally speaking, one expects that the turning point between the high-$Q^2$ and low-$Q^2$ regimes is around the scale of chiral symmetry breaking, $\Lambda_{\chi} \approx 1 ~ GeV$. As $Q^2$ decreases around and below $\Lambda_{\chi}^2$, one expects that the {\em inelastic} coupling of the incoming virtual boson with the $CQ$ becomes less and less important, while the {\em elastic} coupling of the incoming virtual boson with the $CQ$ becomes more and more important. As discussed in Ref.~\cite{PSR} the $Q^2$-range of applicability of the generalized two-stage model is qualitatively given by $\Lambda_{QCD}^2 \lsim Q^2 \lsim \Lambda_{\chi}^2$, i.e. $0.1 \div 0.2 \lsim Q^2 ~ (GeV/c)^2 \lsim 1 \div 2$. 

\indent The dominance of the elastic coupling at the $CQ$ level cannot hold at each $x$ value, but in a local duality sense (a $CQ$-hadron duality), which can be translated in the space of moments into the following (approximate) equivalence
 \be
    M_n^{(T)}(Q^2) \simeq M_n^{(T, dual)}(Q^2)
    \label{eq:dualT}
 \ee
where $M_n^{(T, dual)}(Q^2)$ are the moments evaluated within the generalized two-stage model. Equation (\ref{eq:dualT}) is expected to hold for low values of the order $n$, except $n = 2$. Such a limitation arises from the fact that as $n$ increases the moment $M_n^H(Q^2)$ is more and more sensitive to the rapidly varying bumps of the resonances. Therefore Eq.~(\ref{eq:dualT}) cannot hold at very large values of $n$ (see Refs.~\cite{RGP,duality,SIM00} for the case of the parton-hadron Bloom-Gilman duality \cite{BG}). At the same time it should be pointed out that the dual relation (\ref{eq:dualT}) is expected to hold only for $n > 2$, because the second moment $M_2^{(T)}(Q^2)$ is significantly affected by the low-$x$ region where the concept of valence dominance may become unreliable.

\indent Within the generalized two-stage model the {\em dual} moments $M_n^{(T, dual)}(Q^2)$ can be written in the following factorized form
 \be
     M_n^{(T, dual)}(Q^2) = [F(Q^2)]^2 \cdot \overline{M}_n^{(T)}(Q^2)
     \label{eq:MTndual}
  \ee
where $\overline{M}_n^{(T)}(Q^2)$ describes the effect of the internal $CQ$ motion inside the hadron on the moment of order $n$, and $F(Q^2)$ is the $CQ$ elastic form factor. The latter is defined as \cite{PSR}
 \be
     [F(Q^2)]^2 \equiv {1 \over \sum_j e_j^2} ~ \sum_j { [G_E^j(Q^2)]^2 
     + \tau [G_M^j(Q^2)]^2 \over 1 + \tau } = {1 \over \sum_j e_j^2} ~ 
     \sum_j [f_1^j(Q^2)]^2 + \tau ~ [f_2^j(Q^2)]^2
     \label{eq:FT_CQ}
 \ee
where $f_{1(2)}^j(Q^2)$ and $G_{E(M)}^j(Q^2)$ represent the Dirac(Pauli) and electric(magnetic) Sachs form factors of the $j$-th $CQ$, respectively, and $\tau \equiv Q^2 / 4m^2$ with $m$ being the $CQ$ mass.

\indent If one has a reasonable model for the $CQ$ momentum distributions in the hadron, the moments $\overline{M}_n^{(T)}(Q^2)$ can be estimated and therefore the following ratio
 \be
    R_n^{(T)}(Q^2) \equiv M_n^{(T)}(Q^2) ~ / ~ \overline{M}_n^{(T)}(Q^2)
    \label{eq:ratio}
 \ee
can be constructed starting from the experimental moments $M_n^{(T)}(Q^2)$ [Eq.~(\ref{eq:MTn})]. Within the generalized two-stage model the ratio $R_n^{(T)}(Q^2)$ is expected to depend only on $Q^2$, i.e. it becomes independent of the order $n$ (as well as on the specific hadron), viz. 
 \be
    R_n^{(T)}(Q^2) \simeq [F(Q^2)]^2
    \label{eq:scaling}
 \ee
The scaling function, given by the r.h.s. of Eq.~(\ref{eq:scaling}), is directly the square of the $CQ$ form factor, i.e. the form factor of a confined object.

\indent In Ref.~\cite{PSR} it has been shown that the data of Ref.~\cite{CLAS} manifest a clear tendency to the scaling property (\ref{eq:scaling}) even assuming no internal motion of the $CQ$'s inside the proton, which represents a very simplified and rough model for the $CQ$ momentum distribution. In this case the $CQ$'s share exactly ($1/3$) of the light-front ($LF$) proton momentum. Explicitly one gets \cite{PSR}: $\overline{M}_n^{(T)}(Q^2) \to (1/3)^{n-1}$. Though simple such an hypothesis explains very well the spread of about one order of magnitude between the experimental moments of order $n$ and $(n + 2)$. In our opinion this is an important result (almost a pure experimental result) because it is obtained with a very simple hypothesis about the $CQ$ motion in the proton. 

\indent The effect of the internal $CQ$ motion was investigated and found to play an important role. The final result of the generalized two-stage model is \cite{PSR}
 \be
    \overline{M}_n^{(T)}(Q^2) \equiv \int_0^{\xi^*} d\xi ~ {r (1 + r) \over 2} 
    {\xi^{n+1} \over x^3} {3 + 3 (n + 1) r + n (n + 2) r^2 \over (n + 2) (n 
    + 3)} ~ \xi \overline{f}_2^{TM}(\xi, Q^2) ~ F_{thr}(W) ~~
    \label{eq:MTtheor}
 \ee
where $F_{thr}(W)$ is a factor ensuring the correct behavior at the physical pion threshold, chosen in Ref.~\cite{PSR} to be of the following (parameter-free) form 
 \be
    F_{thr}(W) = \sqrt{1 - \left( {M + m_{\pi} \over W} \right)^2}
    \label{eq:threshold}
 \ee
where $m_{\pi}$($M$) is the pion(nucleon) mass and $W$ is the invariant produced mass [$W^2 = M^2 + Q^2 (1 - x) / x$].

\indent In Eq.~(\ref{eq:MTtheor}) the target-mass-corrected $LF$ momentum distribution $\overline{f}_2^{TM}(\xi, Q^2)$ is given explicitly by
 \be
    \overline{f}_2^{TM}(\xi, Q^2) = {1 \over r^3} \left[ {x^2 \over \xi^2} 
    \overline{f}^p(\xi) + T(\xi, Q^2) \right]
    \label{eq:f2TM}
 \ee
 with
 \be
    T(\xi, Q^2) = {6 M^2 \over Q^2} {x^3 \over r} \int_{\xi}^{\xi^*} d\xi' 
    {\overline{f}^p(\xi') \over \xi' \xi} + {12 M^4 \over Q^4} {x^4 \over 
    r^2} \int_{\xi}^{\xi^*} d\xi' {\overline{f}^p(\xi') \over \xi' \xi} (\xi' 
    - \xi)
    \label{eq:TM}
 \ee
where $x = \xi / (1 - M^2 \xi^2 / Q^2)$ and $\xi^* \equiv \mbox{min}(1, Q / M)$ is the maximum allowed value of the Nachtmann variable $\xi$ (cf. Ref.~\cite{SIM00}). In Eqs.~(\ref{eq:f2TM}-\ref{eq:TM}) the quantity $\overline{f}^p(x)$ is the limiting value of the function $\overline{f}_2^{TM}(\xi, Q^2)$ for $Q^2 \to \infty$ and it represents the probability to find a $CQ$ in the proton with $LF$ momentum fraction $x$. Such a distribution can be estimated assuming an $SU(6)$-symmetric (canonical) wave function with a gaussian ansatz for its radial part. Thus, the target-mass-corrected $LF$ momentum distribution $\overline{f}_2^{TM}(\xi, Q^2)$ used in Ref.~\cite{PSR} has only one model parameter, namely $\beta / m$, where $m$ is the $CQ$ mass and $\beta$ corresponds to the r.m.s. value of the $CQ$ transverse momentum in the proton.

\indent The analysis carried out in Ref.~\cite{PSR} shows that by means of the theoretical moments (\ref{eq:MTtheor}) the scaling property (\ref{eq:scaling}) is well satisfied by the $CLAS$ data for $n = 4, 6, 8$ with the expected exception of $n = 2$. The corresponding scaling function, which represents the square of the $CQ$ elastic form factor $F(Q^2)$, can be interpolated using the square of a monopole ansatz, namely
 \be
     F(Q^2) \simeq 1 / (1 + r_Q^2 ~ Q^2 / 6)
     \label{eq:monopole}
 \ee
The precise value of the $CQ$ size $r_Q$ depends on the specific value of the parameter $\beta / m$ used for the proton wave function and also on the specific shape adopted for the threshold factor $F_{thr}(W)$. Such a dependence has been thoroughly investigated in Ref.~\cite{PSR} and the final result is that a safe estimate of the $CQ$ size $r_Q$ is between $\approx 0.2$ and $\approx 0.3 ~ fm$.

\indent An important consistency check was considered in Ref.~\cite{PSR}, namely: the $CQ$ form factor extracted from the scaling function and the model used for the nucleon wave function should be consistent with the elastic nucleon data. The nucleon elastic form factors were calculated adopting the covariant $LF$ approach of Ref.~\cite{nucleon}, which is formulated at $q^+ = 0$ in order to properly suppress the pair creation process \cite{Zgraph}. Moreover, the one-body approximation for the electromagnetic current operator was considered including both Dirac and Pauli constituent form factors. In Ref.~\cite{PSR} the following simple ans\"atze were chosen:
 \be
     f_1^j(Q^2) & = & e_j / (1 + r_{1 Q}^2 ~ Q^2 / 6) ~ , \nonumber \\[2mm]
     f_2^j(Q^2) & = & \kappa_j / (1 + r_{2 Q}^2 ~ Q^2 / 12)^2 
     \label{eq:CQff}
 \ee
with $r_{1 Q} = r_{2 Q} = r_Q$. The values of the $CQ$ anomalous magnetic moments, $\kappa_U$ and $\kappa_D$, were fixed by the requirement of reproducing the experimental values of proton and neutron magnetic moments, obtaining $\kappa_U = - 0.064$ and $\kappa_D = 0.017$. It turned out \cite{PSR} that the results of the calculation of the nucleon elastic form factors slightly overestimate the data and a better consistency can be reached through slight variations of the parameters of the generalized two-stage model, namely $r_Q$ and $\beta / m$. For instance, a nice agreement with the elastic nucleon data can be recovered simply by increasing the $CQ$ size up to $r_Q = 0.33 ~ fm$.

\indent Before closing this Section, we point out that Eqs.~(\ref{eq:FT_CQ}) and (\ref{eq:monopole}) are not in contradiction with Eq.~(\ref{eq:CQff}). Indeed, as it can be easily checked by explicit calculations, the form factor $F(Q^2)$ is only slightly sensitive to the presence of the Pauli term $f_2^j(Q^2)$, because of the dipole form assumed [see the second equality in Eq.~(\ref{eq:CQff})] and of the smallness of the $CQ$ anomalous magnetic moments. However a Pauli term is expected in non-perturbative models (cf., e.g., Ref.~\cite{instantons}). Therefore we need to analyze a channel different from the transverse one in order to obtain information on $f_2^j(Q^2)$. This chance may be offered by the longitudinal channel as we are going to explain in the next Section.

\section{Longitudinal channel \label{sec:longitudinal}}

\indent Within the generalized two-stage model of Ref.~\cite{PSR} the {\em dual} (transverse) proton structure function, $F_2^{(dual)}(\xi, Q^2)$, is given by
 \be
     F_2^{(dual)}(\xi, Q^2) = \left[ F(Q^2) \right]^2 ~ F_{thr}(W) ~ {\xi 
     \over r^3} \left[ {x^2 \over \xi^2} \overline{f}^p(\xi) + T(\xi, Q^2) 
     \right]
     \label{eq:F2dual}
 \ee
where we have made an explicit use of Eq.~(\ref{eq:f2TM}) and $T(\xi, Q^2)$ is given by Eq.~(\ref{eq:TM}). Using the formalism of target-mass corrections it is straightforward to write down the {\em dual} structure function $F_1^{(dual)}(\xi, Q^2)$; one gets
 \be
     F_1^{(dual)}(\xi, Q^2) = \left[ F(Q^2) \right]^2 ~ F_{thr}(W) ~ 
     {x ~ \xi \over 2r} \left[ {1 + 1 / \tau \over 1 + R_Q} 
     {\overline{f}^p(\xi) \over \xi^2} + {1 \over 3 x^2} T(\xi, Q^2) \right]
     \label{eq:F1dual}
 \ee
where $\tau = Q^2 / 4 m^2$ and 
 \be
     R_Q = {1 \over \tau} {\sum_j \left[ G_E^j(Q^2) \right]^2 \over \sum_j 
     \left[ G_M^j(Q^2) \right]^2}
     \label{eq:RLTQ}
 \ee
represents the longitudinal to transverse ($L/T)$ ratio at the $CQ$ level. Therefore, the {\em dual} longitudinal proton structure function (\ref{eq:FLp}) is
 \be
    F_L^{(dual)}(\xi, Q^2) = \left[ F(Q^2) \right]^2 ~ F_{thr}(W) ~ \xi ~ 
    \left[ {2 \over 3r} T(\xi, Q^2) + {x^2 \over r ~ \xi^2} 
    \overline{f}^p(\xi) {R_Q - 1 / \tau \over 1 + R_Q} \right] ~ .
    \label{eq:FLdual}
 \ee
Thus, within the generalized two-stage model the longitudinal structure function receives two contributions: the first one is directly generated by the target mass corrections $T(\xi, Q^2)$ and depends on the form factor $F(Q^2)$, which is almost insensitive to the $CQ$ Pauli form factor; the second one is instead directly related to the $L/T$ ratio $R_Q$ at the $CQ$ level. In the limit of large $Q^2$ both the first and the second contribution drop down and for $Q^2 \to \infty$ the function $F_L^{(dual)}(\xi, Q^2)$ goes to zero, satisfying the Callan-Gross relation.

\indent It is important to note that if $f_2^j(Q^2) = 0$ then $G_E^j(Q^2) = G_M^j(Q^2)$. Thus $R_Q$ becomes equal to $(1 / \tau)$ and the second term in the r.h.s of Eq.~(\ref{eq:FLdual}) vanishes. Therefore, $R_Q$ must be at least proportional to the $CQ$ Pauli form factor and consequently we expect the second term of the r.h.s. of Eq.~(\ref{eq:FLdual}) to be sensitive to $f_2^j(Q^2)$.

\begin{figure}[htb]

\centerline{\epsfig{file=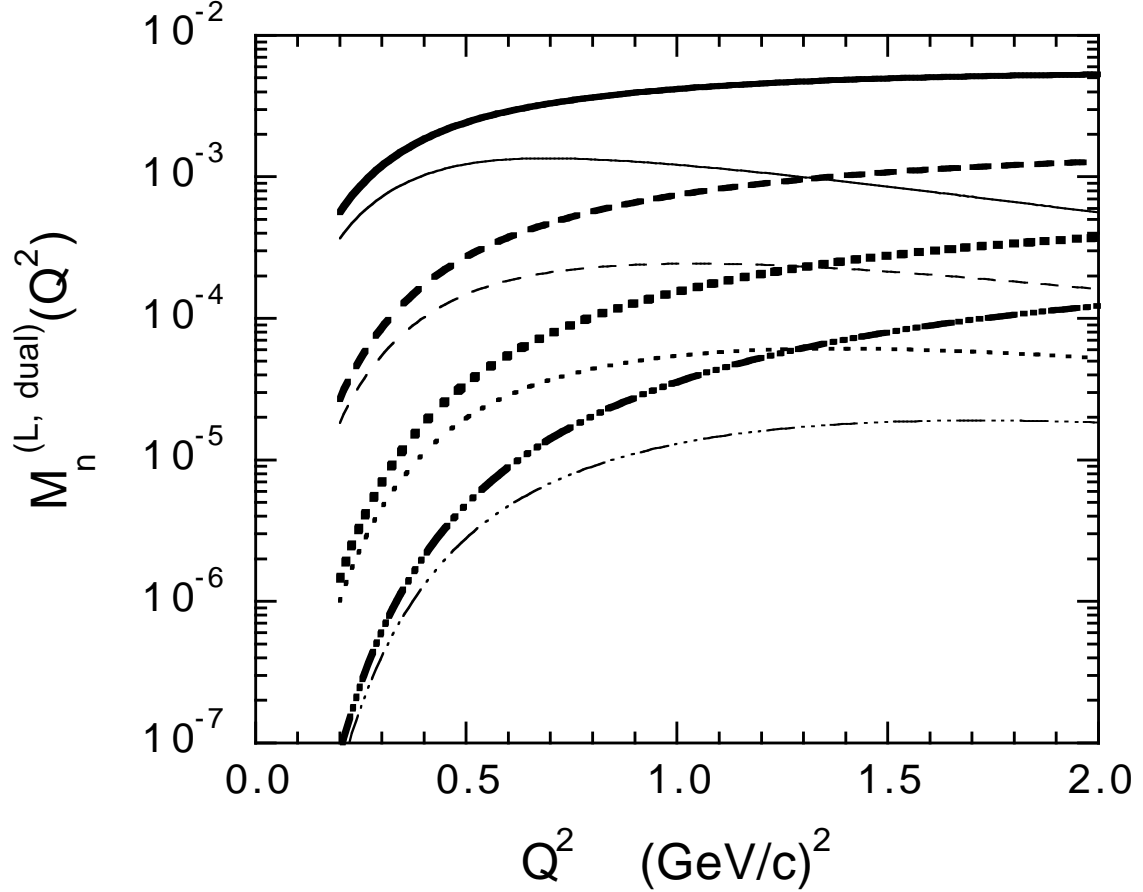,width=15cm}}

\caption{\label{fig:comparison} \small \em Dual longitudinal moments [Eq.~(\ref{eq:MLtheor})], calculated using the structure functions $F_L^{(dual)}$ [Eq.~(\ref{eq:FLdual})] and $F_2^{(dual)}$ [Eq.~(\ref{eq:F2dual})], versus $Q^2$. The $CQ$ Dirac form factor is given by the monopole ansatz $f_1^j(Q^2) = e_j / (1 + r_Q^2 ~ Q^2 / 6)$ with $r_Q = 0.21 ~ fm$ \cite{PSR}. Thin lines correspond to $f_2^j(Q^2) = 0$, while thick lines are the results obtained with $f_2^j(Q^2) = k_j / (1 + r_Q^2 ~ Q^2 / 12)^2$ with $r_Q = 0.21 ~ fm$. The values of the $CQ$ anomalous magnetic moments are taken from Ref.~\cite{PSR}, namely: $\kappa_U = - 0.064$ and $\kappa_D = 0.017$. Solid, dashed, dotted and triple-dotted-dashed curves correspond to $n = 4, 6, 8, 10$, respectively.}

\end{figure}

\indent According to the generalized two-stage model we expect that for $0.1 \div 0.2 \lsim Q^2 ~ (GeV/c)^2 \lsim 1 \div 2$ the (inelastic) Nachtmann longitudinal moments (\ref{eq:MLn}) are dominated for low values of $n$ (but $n = 2$) by the {\em dual} longitudinal moments, viz.
 \be
    M_n^{(L)}(Q^2) \simeq M_n^{(L, dual)}(Q^2)
    \label{eq:dualL}
 \ee
where
 \be
    M_n^{(L, dual)}(Q^2) & \equiv & \int_0^{\xi^*} d\xi ~ {r (1 + r) \over 2} 
    {\xi^{n+1} \over x^3} \left[ F_L^{(dual)}(\xi, Q^2) \right. 
    \nonumber \\[2mm]
    & + & \left. {4 M^2 x^2 \over Q^2} F_2^{(dual)}(\xi, Q^2) {(n + 1) ~ \xi / 
    x - 2 (n + 2) \over (n + 2) (n + 3) } \right] 
    \label{eq:MLtheor}
 \ee
We have calculated the {\em dual} moments (\ref{eq:MLtheor}) for $n = 4, 6, 8, 10$ using the structure functions $F_L^{(dual)}$ and $F_2^{(dual)}$, given respectively by Eqs.~(\ref{eq:FLdual}) and (\ref{eq:F2dual}). The same gaussian ansatz adopted in Ref.~\cite{PSR} (with $\beta = 0.3 ~ GeV$ and $m = 0.25 ~ GeV$) is used for the proton wave function and the expressions (\ref{eq:CQff}) with $r_{1 Q} = r_{2 Q} = r_Q$ are adopted for the $CQ$ form factors. The results obtained with and without the $CQ$ Pauli term are reported in Fig.~\ref{fig:comparison}. It can clearly be seen that above $Q^2 \approx 0.5 ~ (GeV/c)^2$ the longitudinal moments are sharply sensitive to the presence of the $CQ$ Pauli form factor, i.e. to the second term in the r.h.s of Eq.~(\ref{eq:FLdual}).

\begin{figure}[htb]

\centerline{\epsfig{file=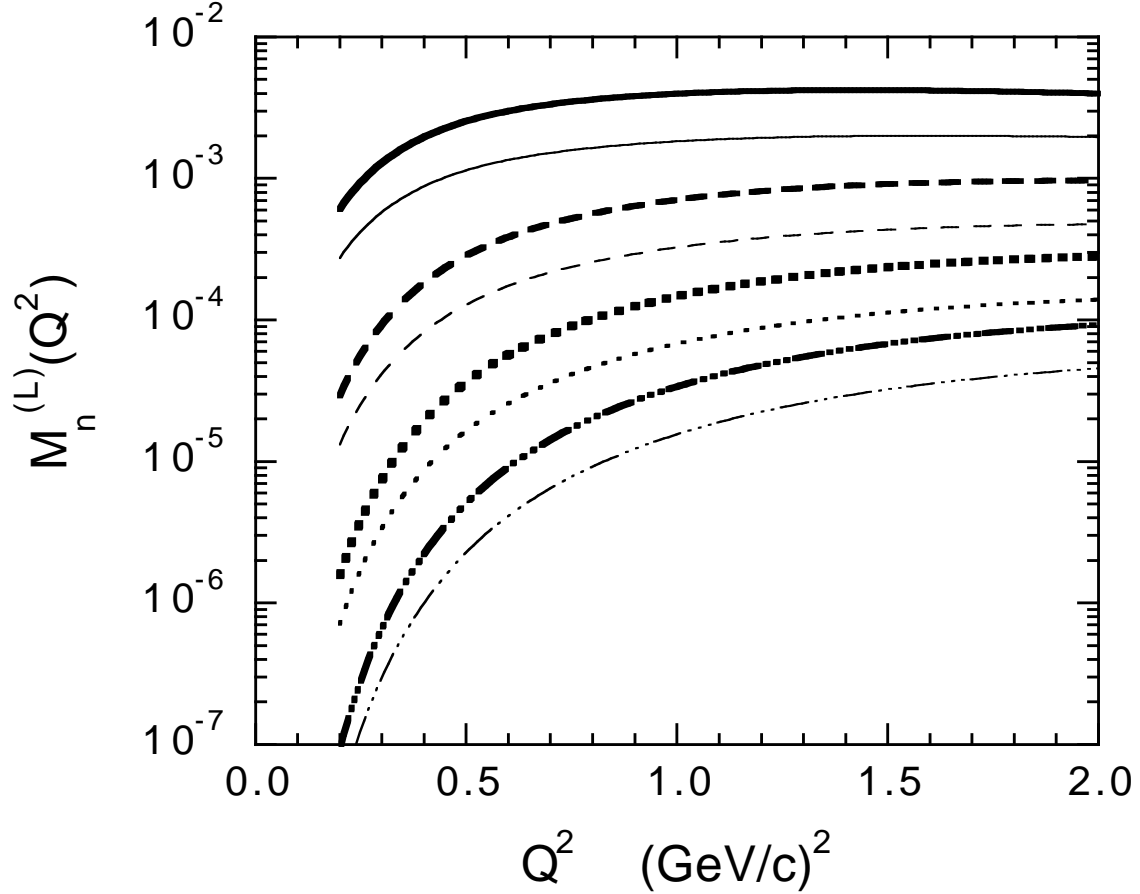,width=15cm}}

\caption{\label{fig:prediction} \small \em Inelastic contribution to the Nachtmann longitudinal moments (\ref{eq:MLn}) evaluated through the dual moments (\ref{eq:MLtheor}) using the structure functions $F_L^{(dual)}$ [Eq.~(\ref{eq:FLdual})] and $F_2^{(dual)}$ [Eq.~(\ref{eq:F2dual})], versus $Q^2$. The $CQ$ Dirac and Pauli form factors are given by Eq.~(\ref{eq:CQff}) with $\kappa_U = - 0.064$ and $\kappa_D = 0.017$. Solid, dashed, dotted and triple-dotted-dashed curves correspond to $n = 4, 6, 8, 10$, respectively. The difference between the thin and thick lines represent the theoretical uncertainties of our predictions as explained in the text.}

\end{figure}

\indent We present now our predictions for the (inelastic) longitudinal moments (\ref{eq:MLn}) through the {\em dual} longitudinal moments (\ref{eq:MLtheor}) for $n \geq 4$. We stress that our model parameters have the same values found in Ref.~\cite{PSR} and that no further parameters are introduced. In particular, the $CQ$ form factors are given by Eq.~(\ref{eq:CQff}) and we start from the choice $r_{1 Q} = r_{2 Q} = r_Q = 0.33 ~ fm$, which as already pointed out provide the best reproduction of nucleon elastic data. Since there is no compelling reason to assume the same $CQ$ size both in Dirac and in Pauli form factors, we have varied independently the parameters $r_{1 Q}$ and $r_{2 Q}$ in the range from $0.2$ to $0.4 ~ fm$. The spread of the results obtained in this way for each longitudinal moment, is represented by the difference between the thin and thick lines in Fig.~\ref{fig:prediction}, which therefore yields our estimate of the theoretical uncertainties. 

\section{Conclusions \label{sec:conclusions}}

\indent In conclusion we have applied the generalized two-stage model of Ref.~\cite{PSR} to the calculation of the (inelastic) Nachtmann moments of the longitudinal structure function of the proton at low momentum transfer, i.e. $0.2 \lsim Q^2 ~ (GeV/c)^2 \lsim 2$. It has been shown that in the longitudinal channel the Pauli form factor of the constituent quarks play an important role for $Q^2 \gsim 0.5 ~ (GeV/c)^2$.

\indent Our predictions are presented in Fig.~\ref{fig:prediction}, including also an estimate of the theoretical uncertainties. We point out that our model parameters have the same values found in Ref.~\cite{PSR} and no further parameters are introduced. 

\indent Our predictions may be tested against the forthcoming results of the Jefferson Lab experiment $E94110$ \cite{HALLC}. A positive comparison with the new data will provide a strong confirmation of the generalized two-stage model of Ref.~\cite{PSR} and a compelling evidence that constituent quarks are intermediate substructures between the hadrons and the current quarks and gluons of $QCD$.

\section*{Acknowledgments}

\indent The author gratefully acknowledges R.~Petronzio and G.~Ricco for many valuable comments.

\end{document}